\documentclass[12pt,epsbox]{article}
\usepackage{color}
\usepackage[dvips]{graphicx}
\usepackage{amsmath}

\baselineskip=\normalbaselineskip

\newcommand{\resection}[1]
 {\setcounter{equation}{0}\section{\large{#1}}}

\renewcommand{\thefootnote}{\fnsymbol{footnote}}
\newcommand{\bel}[1]{\begin{equation}\label{#1}}
\newcommand{\bal}[1]{\begin{eqnarray}\label{#1}}

\newcommand{\be}{\begin{equation}}
\newcommand{\ee}{\end{equation}}
\newcommand{\ba}{\begin{eqnarray}}
\newcommand{\ea}{\end{eqnarray}}
\newcommand{\nn}{\nonumber \\}

\newcommand{\n}{\nonumber}

\newcommand{\bR}{{\bf R}}

\newcommand{\eq}[1]{(\ref{#1})}

\renewcommand{\bR}{\mbox{\boldmath $R$}}

\newcommand{\Gammaeff}{\Gamma_{\!{\rm eff}}}
\begin{document}
\setcounter{page}{0}
\begin{flushright}
\parbox{40mm}{%
KUNS-1832 \\
YITP-03-12 \\
{\tt hep-th/0304012} \\
April 2003}

\end{flushright}

\vfill

%%%%%%%%%%%%%%%%%%%%%%%%%%%%%%%%%%%%%%%%%%%%%%%%%%%%%
%% Title
%%%%%%%%%%%%%%%%%%%%%%%%%%%%%%%%%%%%%%%%%%%%%%%%%%%%
\begin{center}
{\large{\bf 
Holographic entropy bound 
in two-dimensional gravity
}}
\end{center}

\vfill

%%%%%%%%%%%%%%%%%%%%%%%%%%%%%%
%% author
%%%%%%%%%%%%%%%%%%%%%%%%%%%%%%
\begin{center}
{\sc Masafumi Fukuma}$^{1}$\footnote%
{E-mail: {\tt fukuma@gauge.scphys.kyoto-u.ac.jp}},  
{\sc Akitsugu Miwa}$^{1}$\footnote%
{E-mail: {\tt akitsugu@gauge.scphys.kyoto-u.ac.jp}} and 
{\sc Kazuyoshi Takahashi}$^{2}$\footnote%
{E-mail: {\tt kazuyosi@yukawa.kyoto-u.ac.jp}} \\[2em]
$^1${\sl Department of Physics, Kyoto University, Kyoto 606-8502, Japan} \\
$^2${\sl Yukawa Institute for Theoretical Physics, 
      Kyoto University, Kyoto 606-8502, Japan } \\

\end{center}

\vfill
%%%%%%%%%%%%%%%%%%%%%%%%%%%%%%%%%
%% abstract
%%%%%%%%%%%%%%%%%%%%%%%%%%%%%%%%%
\begin{center}
ABSTRACT
\end{center}

\begin{quote}

\small{%
Bousso's entropy bound for two-dimensional gravity 
is investigated in the lightcone gauge. 
It is shown that due to the Weyl anomaly, 
the null component of the energy-momentum tensor 
takes a nonvanishing value, 
and thus, combined with the conditions that were recently proposed 
by Bousso, Flanagan and Marolf, 
a holographic entropy bound similar to Bousso's 
is expected to hold in two dimensions. 
A connection of our result to that of Strominger and Thompson 
is also discussed. 
}
\end{quote}
\vfill
%%

%%%%%%%%%%%%%%%%%%%%%%%%%%%%%%%%%%
% Main
%%%%%%%%%%%%%%%%%%%%%%%%%%%%%%%%%%
\renewcommand{\thefootnote}{\arabic{footnote}}
\setcounter{footnote}{0}
\addtocounter{page}{1}
%%%%%%%%%%%%%%%%%%%%%%%%%%%%%%%%%
\newpage

\resection{Introduction}

The ultimate goal of today's high energy physics is 
to unify all the interactions including gravity. 
Although string theory is the most promising candidate 
for such ``theory of everything," 
a constructive definition of string theory is still under development. 
For this purpose, it is relevant to grasp 
the fundamental dynamical variables (even if they are strings) 
that describe gravity systems in a consistent manner 
and also in such a way that all observations which have been made on gravity 
are naturally accounted for. 

The holographic principle \cite{'tHooft:gx}\cite{Susskind:1994vu} 
appears in the developments 
of the study of thermodynamics of black holes 
and various entropy bounds, 
and states that a spacelike or lightlike region of 
a system including gravity can be equivalently described by 
the system on the boundary of the region 
(nice review articles are \cite{Bigatti:1999dp}\cite{Bousso:2002ju}). 
In order to check this principle quantum mechanically, 
we need to have a full quantum theory of gravity, 
but the current status of string theory is 
not at the level to check in a satisfactory manner 
whether the assertion holds or not. 
On the other hand, gravity can be treated fully quantum mechanically 
in two dimensions, 
so that it is worthwhile to study the holographic nature 
in two-dimensional gravity. 

Here we first make a historical review of the holographic principle. 
%%%%%%%%%%%%%%%%%%%%%%%%%%%%%%%%%%%%%%%%%%%%%%%%%%%%%
%In 1973, Bardeen, Cartar and Hawking \cite{Bardeen:gs} proposed 
%that the entropy of 
%a black hole should be equal to a quarter of the horizon area. 
%%%%%%%%%%%%%%%%%%%%%%%%%%%%%%%%%%%%%%%%%%%%%%%%%%%%%
In the early 70th, Bekenstein proposed that the entropy of a black
hole, $S_{\rm BH}$, is proportional 
to the area of the horizon \cite{Bekenstein:1972tm}\cite{Bekenstein:ur}, 
and subsequently, Bardeen, Cartar and Hawking \cite{Bardeen:gs}
and Hawking \cite{Hawking:sw} determined the coefficient to be 
$1/4G_{\rm N}$, {\em i.e.}, $S_{\rm BH} = A/4G_{\rm N}$, 
where $A$ and $G_{\rm N}$ are the area of the horizon and the Newton constant, 
respectively.\footnote{%
We will set $G_{\rm N}=1$ in the following discussions.}
%%%%%%%%%%%%%%%%%%%%%%%%%%%%%%%%%%%%%%%%%%%%%%%%%%%%%
On the basis of their proposal  
and by requiring the generalized second law of thermodynamics 
\cite{Bekenstein:ax} 
(matters' entropy plus black holes' entropy should increase in time), 
Bekenstein \cite{Bekenstein:jp} showed 
that there must be a bound on the entropy of matters 
in the following form: 
\begin{equation}
 S \leq 2\pi ER,
\end{equation}
where $S$, $E$ and $R$ are the entropy, the energy and the linear size  
of the system, respectively. 
This entropy bound is expected to hold for systems with weak gravity. 
On the other hand, 't Hooft \cite{'tHooft:gx} 
and Susskind \cite{Susskind:1994vu} pointed out that, 
for strongly gravitating systems, 
we must use another entropy bound %(Holographic Entropy bound):
\begin{equation}
 S \leq \frac{1}{4}\,A, 
 \label{HOLOBOUND}
\end{equation}
where $A$ is the area of the boundary of a given region. 
%%%%%%%%%%%%%%%%%%%%%%%%%%%%%%%%%%%%%%%%%
%where $A$ is the area of the boundary of the system, 
%and $G_{\rm N}$ is the Newton constant. 
%%%%%%%%%%%%%%%%%%%%%%%%%%%%%%%%%%%%%%%%%
The bound saturates when the region is filled with a black hole 
whose horizon coincides with the boundary. 
This bound \eq{HOLOBOUND} is much more radical than the Bekenstein bound, 
because it asserts that the entropy of gravity system is not extensive 
and should be bounded by area. 
It thus suggests that the fundamental dynamical variables 
including gravity are not ordinary local variables 
but correspond to some quantities living on the boundary. 
This is called the holographic principle. 

Originally, the bound \eq{HOLOBOUND} is assumed to hold for 
the entropy of a spacelike region, 
but there are many counter examples to such spacelike 
entropy bound. 
One such counter example is the flat Friedmann-Robertson-Walker (FRW) universe.
Fischler and Susskind \cite{Fischler:1998st} 
pointed out that if the entropy is estimated for a lightlike region 
(null hypersurface) then the same form of entropy bound does hold 
even for the flat FRW universe. 
Subsequently Bousso made their proposal into a precise form 
\cite{Bousso:1999xy}\cite{Bousso:1999cb},  
introducing the idea of ``lightsheet," 
which is a null hypersurface characterized by 
the condition that the expansion $\theta$ is nonpositive 
along the generator of the null hypersurface. 
\begin{figure}[htbp]
\begin{center}
\resizebox{!}{40mm}{
 \input{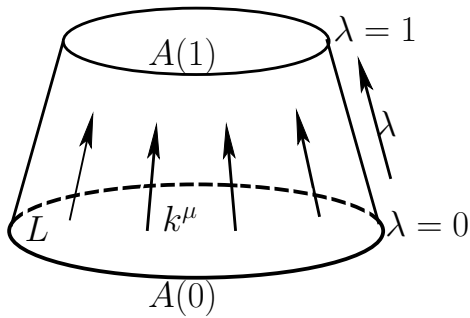}
}
\caption{\footnotesize{
 A null hypersurface $L$ generated by a null vector
 $k=d/d\lambda$ with affine parameter $\lambda$. 
 $L$ is parametrized from $\lambda=0$ to $\lambda=1$. 
 Letting $A(\lambda)$ be the cross-sectional area at $\lambda$, 
 the expansion $\theta$ is defined as 
 $\theta(\lambda)\equiv A'(\lambda)/A(\lambda)$. 
 If the expansion is always non-positive along $\lambda$, 
 $\theta(\lambda)\leq 0$, the null hypersurface $L$ is called 
 (part of) a lightsheet. 
 The Bousso entropy bound states that the entropy $S_L$ on the lightsheet 
 is bounded by the difference of the areas $A(0)$ and $A(1)$ 
 of the boundaries of the lightsheet, 
 $S_L \leq (1/4)\bigl(A(0)-A(1)\bigr)$.}}
\label{lightsheet}
\end{center}
\end{figure}
His entropy bound requires the entropy $S_L$ on the lightsheet $L$ 
to be bounded by the difference of the areas $A(0)$ and $A(1)$ 
of the boundaries of the lightsheet (see fig.\ref{lightsheet}):\footnote{%
Bousso originally proposed the bound as 
$S_L\leq (1/4)\,A(0)$. 
The entropy bound in the form of eq.\ \eq{bousso} was first presented  
by Flanagan, Marolf and Wald \cite{Flanagan:1999jp}.}
\begin{equation}
 S_L \leq \frac{1}{4}\bigl(A(0)-A(1)\bigr).
 \label{bousso}
\end{equation}
This bound has not had serious counter examples so far 
and is believed to hold for any classical gravity.

A derivation of the Bousso bound \eq{bousso} was given by 
Flanagan, Marolf and Wald (FMW) \cite{Flanagan:1999jp}, 
who assumed a condition that is essentially equivalent to 
the Bekenstein bound on a null hypersurface: 
\ba
 s\equiv -s^\mu\,k_\mu\leq 2\pi\,(1-\lambda)\,
  T_{\mu\nu}\,k^\mu k^\nu.
 \label{FMW}
\ea
Here $s^\mu$ and $s $ are the entropy current and the entropy density 
over a lightsheet $L$, respectively, 
and the null vector $k\equiv d/d\lambda$ is the generator of the lightsheet 
with affine parameter $\lambda$ 
(for detailed definitions, see the next section). 

The FMW condition \eq{FMW} %is expected to hold for ordinary systems, but 
is not given in a completely local form, 
as can be seen, for example, from the presence of $\lambda$ in the expression. 
Recently, Bousso, Flanagan and Marolf (BFM) \cite{BFM:2003} 
gave another condition which is completely local: 
\ba
 s'\equiv \frac{d s}{d\lambda}\leq 2\pi\,T_{\mu\nu}\,k^\mu k^\nu. 
 \label{BFM}
\ea
This is essentially a differential form of the FMW condition. 
As discussed in the next section, 
this condition leads to the Bousso bound 
in $(n+2)$ dimensions ($n\geq1$). 
The essential step in the derivation there 
is to use the identity 
\ba
 T_{\mu\nu}\,k^\mu k^\nu(\lambda) 
  =-\frac{n}{8\pi}\,\frac{G''(\lambda)}{G(\lambda)}
  -\frac{1}{8\pi}\,\sigma_{\mu\nu}\,\sigma^{\mu\nu},
 \label{ident}
\ea
which can be proven by the Raychaudhuri equation 
(see the next section). 
Here $\sigma_{\mu\nu}$ is the shear tensor, 
and 
\ba
 G(\lambda)\equiv \exp\Bigl[\,\frac{1}{n}
  \int_0^\lambda d\bar{\lambda}\,\theta(\bar{\lambda})\Bigr].
\ea

In the present article, we discuss two-dimensional gravity $(n=0$) 
which is coupled to conformal matters of central charge $c$, 
treating gravity as a background for quantum matters. 
If we set $n=0$, the right-hand side of eq.\ \eq{ident} vanishes 
since the shear tensor does not exist in two dimensions, 
so that eqs.\ \eq{BFM} and \eq{ident} do not seem to give 
a physically meaningful bound on the entropy. 
The main purpose of this letter is to show 
that due to the Weyl anomaly, 
an equality similar to eq.\ \eq{ident} holds for two dimensions: 
\ba
 T_{\mu\nu}\,k^\mu k^\nu(\lambda) 
  =-4\beta\,\frac{\bar{G}''(\lambda)}{\bar{G}(\lambda)},
 \label{ident2}
\ea
where the coefficient is related to the central charge of matters 
as $\beta=(c-26)/48\pi$, 
and $\bar{G}(\lambda)$ is related to an effective ``area" 
of point at $\lambda$. 
This equality allows us to prove a nontrivial bound on the entropy. 
In order to discuss fully quantum aspects of the Bousso bound, 
we further need to quantize gravity. 
This is now under investigation and will be reported elsewhere.

This letter is organized as follows. 
In section 2, we first review the BFM conditions 
in $(n+2)$-dimensional space-time, 
which leads to a classical bound on the entropy over a lightsheet. 
The derivation does not work in two dimensions 
when gravity systems are treated classically. 
However, in section 3, we show that the same manner of derivation 
is also possible in two dimensions 
when the Weyl anomaly is taken into account correctly. 
In section 4, we compare our result with the energy-momentum tensor 
of two-dimensional dilatonic gravity, 
which can be obtained 
by compactifying an $(n+2)$-dimensional space-time 
on $n$-dimensional sphere, $S^n$. 
This section was inspired by work of Strominger and Thompson 
\cite{Strominger:2003br}. 
Section 5 is devoted to conclusion and outlook. 

%%%%%%%%%%%%%%%%%%%%%%%%%%%%%%%%%
\resection{BFM conditions}

Let $(M,g)$ be an $(n+2)$-dimensional space-time 
with metric $g=(g_{\mu\nu})$. 
Let $L$ be a null hypersurface which is generated 
by a null vector $k$ with affine parameter $\lambda$,\footnote{
$\lambda$ is called an affine parameter 
if $k=d/d\lambda=k^\mu \partial_\mu$ satisfies the geodesic equation 
of the form $k^\nu\nabla_\nu k^\mu=0$. 
}
\ba
 k=\frac{d}{d\lambda}. 
\ea
We assume that $L$ is parametrized 
from $\lambda=0$ to $\lambda=1$ with two boundaries. 
Denoting the cross-sectional area at $\lambda$ by $A(\lambda)$, 
and its ratio to $A(0)$ by $a(\lambda)$;
\ba
 a(\lambda)\equiv\frac{A(\lambda)}{A(0)}, 
\ea
we introduce the expansion $\theta(\lambda)$ as 
\ba
 \theta(\lambda)\equiv\frac{1}{A(\lambda)}\,\frac{dA(\lambda)}{d\lambda}
  =\frac{1}{a(\lambda)}\,\frac{d a(\lambda)}{d\lambda}.
 \label{theta}
\ea
Then $L$ is called (part of) a lightsheet if the expansion $\theta$ 
is always non-positive along $\lambda$, $\theta(\lambda)\leq0$. 

By denoting the entropy current by $s^\mu$ 
and the entropy density by $s\equiv -s_\mu k^\mu$ 
on the lightsheet $L$, 
Bousso, Flanagan and Marolf \cite{BFM:2003} showed that the Bousso bound 
\eq{bousso} can be derived if the following two conditions are satisfied: 
\begin{align}
 {\rm (i)}&~~s'(\lambda)\equiv \frac{d s(\lambda)}{d\lambda}
  \leq 2\pi\,T_{\mu\nu}\,k^\mu\,k^\nu 
  ~~~~~~~~~~~~~~~~~~~~~~~~~~~~~~~~~~~~~~~~~~~~~~~~~~\\
 {\rm (ii)}&~~s(0)\leq -\frac{1}{4}\,a'(0)
  =-\frac{1}{4}\,\frac{A'(0)}{A(0)}. 
\end{align} 
The first condition is a differential form 
of the FMW condition, which is an analogue of the Bekenstein bound 
and thus is supposed to hold for any normal matters. 
The second one is an initial condition, whose physical meaning 
is described in \cite{Strominger:2003br}. 

We here demonstrate that the above two conditions actually lead to 
the Bousso bound \eq{bousso} in arbitrary $(n+2)$ dimensions ($n\geq1$). 
We first introduce the function $G(\lambda)$ as 
\ba
 G(\lambda)\equiv \bigl[ a(\lambda)\bigr]^{1/n}. 
\ea
Then the Raychaudhuri equation\footnote{
Here, $\sigma_{\mu\nu}$ and $\omega_{\mu\nu}$ 
are the shear and the twist tensors, respectively 
\cite{Wald:tex}. 
The latter vanishes when the vector $k^\mu$ generates 
a family of null hypersurfaces, 
to which $k^\mu$ is normal (and tangent).} 
\ba
 \frac{d\theta}{d\lambda}=-\frac{1}{n}\,\theta^2
  -\sigma_{\mu\nu}\,\sigma^{\mu\nu}+\omega_{\mu\nu}\,\omega^{\mu\nu}
  -8\pi\,T_{\mu\nu}\,k^\mu k^\nu
\ea
gives an inequality on $T_{\mu\nu}\,k^\mu k^\nu$ as
\ba
 T_{\mu\nu}\,k^\mu k^\nu(\lambda) 
  =-\frac{n}{8\pi}\,\frac{G''(\lambda)}{G(\lambda)}
  -\frac{1}{8\pi}\,\sigma_{\mu\nu}\,\sigma^{\mu\nu}
  \leq -\frac{n}{8\pi}\,\frac{G''(\lambda)}{G(\lambda)}
  \leq -\frac{n}{8\pi}\,
   \Bigl(\frac{G'(\lambda)}{G(\lambda)}\Bigr)'.
\ea
We thus have the following inequality on the entropy density: 
\begin{align}
 s(\lambda)&=\int_0^\lambda d\bar{\lambda}\,s'(\bar{\lambda})+s(0)\nn
 &\leq 2\pi\,\int_0^\lambda d\bar{\lambda}\,
  T_{\mu\nu}\,k^\mu k^\nu(\bar{ \lambda})+s(0)\nn
 &\leq -\frac{n}{4}\,\frac{G'(\lambda)}{G(\lambda)}
  +\frac{n}{4}\,\frac{G'(0)}{G(0)}+s(0)\nn
 &\leq -\frac{n}{4}\,\frac{G'(\lambda)}{G(\lambda)}\qquad
  \left(\frac{n}{4}\,\frac{G'(0)}{G(0)}
  =\frac{1}{4}\,a'(0)\leq -s(0)\right), 
\end{align}
which gives a bound on the entropy over the lightsheet $L$ 
($0\leq\lambda\leq 1$) as 
\begin{align}
 S_L &
% \equiv A(0)\,\int_0^1 d\lambda\,s(\lambda)\,a(\lambda) \nn
% &
=A(0)\,\int_0^1 d\lambda\,s(\lambda)\,\bigl[G(\lambda)\bigr]^n\nn
 &\leq - \frac{n}{4}\,A(0)\,\int_0^1 
  d\lambda\,G'(\lambda)\,\bigl[G(\lambda)\bigr]^{n-1}\nn
 &=- \frac{1}{4}\,A(0)\,\bigl[a(\lambda)\bigr]_0^1\nn
 &=\frac{1}{4}\bigl(A(0)-A(1)\bigr).
\end{align}

%%%%%%%%%%%%%%%%%%%%%%%%%%%%%%%%%%%%%%%%%%%%%%%%%%%%%
\resection{2D gravity in the lightcone gauge}

In the preceding section, we have shown that 
the classical identity  
\ba
 T_{\mu\nu}\,k^\mu k^\nu(\lambda) 
  = -\frac{n}{8\pi}\,\frac{G''(\lambda)}{G(\lambda)}
   -\frac{1}{8\pi}\,\sigma_{\mu\nu}\,\sigma^{\mu\nu},
 \label{ident3}
\ea
plays an essential role in deriving the Bousso bound 
in $(n+2)$-dimensional space-time. 
One might think that the equality makes no sense 
in two dimensions, 
since the shear tensor does not exist in two dimensions 
and thus the right-hand side vanishes when we set $n=0$. 
The main purpose of this section is to show that 
if we take into account the Weyl anomaly correctly, 
the right-hand side is rewritten into the desired form 
with a nonvanishing coefficient,  
and that the coefficient is essentially the central charge 
of the conformal matter to which gravity is coupled. 

We consider the effective action $\Gammaeff[g(x)]$ defined as
\ba
 e^{i\Gammaeff[g(x)]}=\int [d\phi(x)]\,e^{i S[\phi(x),g(x)]},
\ea
where $\phi(x)$ stands for a set of conformal matters. 
Assuming that the path integral is regularized in such a way 
that two-dimensional diffeomorphism (${\rm Diff}_2$) is respected, 
the effective action can be calculated 
by integrating the Weyl anomaly equation: 
\ba
 \left\langle T(x) \right\rangle_g\equiv
  \frac{2}{\sqrt{-g}}\,g_{\mu\nu}\,
  \frac{\delta\Gammaeff}{\delta g_{\mu\nu}(x)}
  =\frac{c-26}{24\pi}\,R. 
\ea
The result is \cite{Polyakov:1987zb}
\ba
 \Gammaeff[g]=\frac{\beta}{2}\,\int d^2 x\,\sqrt{-g}\,
  R\,\frac{1}{\nabla^2}\,R
\ea
with\footnote{
Here $-26$ comes from the Jacobian when we reduce 
the measure over $g_{\mu\nu}$ to the measure over $h$ \cite{Polyakov:rd}:
\ba
 \frac{\bigl[dg_{\mu\nu}\bigr]}{{\rm Vol}({\rm Diff}_2)}
  =\bigl[dh\bigr]\times \bigl({\rm Jacobian}\bigr), \n
\ea
although we do not make an integration over $h$ in this article. } 
\ba
 \beta\equiv \frac{c-26}{48\pi}. 
\ea

There are two popular parametrizations (or gauges) of metric; 
one is the conformal gauge and the other is the lightcone gauge. 
Although the former has an advantage in its manifest covariance, 
the lightcone gauge would be more convenient 
in order to analyze holographic behavior over lightsheets. 
We thus write the metric as 
\ba
 ds^2=g_{\mu\nu}(x)\,dx^\mu dx^\nu
  =-dx^+\bigl( dx^- \!+h(x^+,x^-)\,dx^+\bigr). 
\ea
{}For this, $x^+={\rm const}$ gives a null hyper ``surface," 
which is generated by $k\equiv\partial_-=\partial/\partial x^-$ 
with affine parameter $x^-$. 
Then $\Gammaeff[g]=\Gammaeff[h]$ is expressed as \cite{Polyakov:1987zb}
\ba
 \Gammaeff[h]=\beta \int d^2 x \left[
  \frac{\partial_-^2f\,\partial_-\partial_+f}{\bigl(\partial_-f\bigr)^2}
  -\frac{\bigl(\partial_-^2f\bigr)^2\,\partial_+f}
  {\bigl(\partial_-f\bigr)^3} \right]. 
 \label{gamma_h}
\ea
Here we have introduced the function $f(x^+,\,x^-)$ 
through the relation 
\ba
 h(x)=\frac{\partial_+ f(x)}{\partial_- f(x)}.
\ea
{}From eq.\ \eq{gamma_h} the energy-momentum tensor is calculated as 
\begin{align}
 T_{--}&\equiv T_{\mu\nu}\,k^\mu k^\nu \nn
 &= \frac{2}{\sqrt{-g}}\,\frac{\delta \Gammaeff}{\delta g_{\mu\nu}}\,
  k_\mu k_\nu \nn
 &=-4\beta\,\sqrt{\partial_-f}\,\partial_-^2\, \frac{1}{\sqrt{\partial_-f}}.
 \label{Tmm}
\end{align}
This has the same form with the equation \eq{ident3} 
under the identification 
\ba
\begin{array}{ccc}
 \lambda & \leftrightarrow & x^- \\
 G & \leftrightarrow & 1/\sqrt{\partial_-f} \\
 n & \leftrightarrow & (2/3)\,(c-26).
\end{array}
\label{comp}
\ea

Note that for arbitrary function $f$ of $x$, 
the expression $\sqrt{f'(x)}\,\Bigl(1/\sqrt{f'(x)}\Bigr)''$ 
is essentially the Schwarzian differential: 
\ba
 \bigl\{ f,\,x\bigr\}
  \equiv\frac{f'(x)\,f'''(x)-(3/2)\,\bigl(f''(x)\bigr)^2}{
    \bigl(f'(x)\bigr)^2}
  =-2\,\sqrt{f'(x)}\,\Bigl(\frac{1}{\sqrt{f'(x)}}\Bigr)''.
\ea
{}From this, we can easily understand why $T_{--}$ has 
the form \eq{Tmm}. 
In fact, consider the diffeomorphism $F$ defined by
\ba
 F:\,\bigl(x^+,\,x^-\bigr) \,
  \rightarrow\,\bigl(\tilde{x}^+,\,\tilde{x}^-\bigr) 
  =\bigl(x^+,\,f(x^+,x^-)\bigr).
\ea
This is actually a conformal isometry 
from $ds^2=-dx^+\bigl(dx^-\!+h\,dx^+\bigr)$ 
to $d\tilde{s}^2\equiv -d\tilde{x}^+\,d\tilde{x}^-$ 
since 
\begin{align}
 F^{\,\ast}\bigl(d\tilde{s}^2\bigr)
  &=-\,d\tilde{x}^+(x)\,d\tilde{x}^-(x)
  =-\,dx^+\,\Bigl(\bigl(\partial_+f\bigr)\,dx^+ 
    +\bigl(\partial_-f\bigr)\,dx^-\Bigr)\nn
 &=-\,\bigl(\partial_-f\bigr)\,dx^+\,\bigl(dx^-\! +h\,dx^+\bigr)
  =\bigl(\partial_-f\bigr)\,ds^2. 
\end{align}
The energy-momentum tensor vanishes in the $\tilde{x}$ coordinates 
($\widetilde{T}_{--}(\tilde{x})=0$), 
and thus, by using the transformation law of the energy-momentum tensor,  
$T_{--}(x)$ can be calculated as follows: 
\begin{align}
 T_{--}(x)&=\bigl(\partial_-f\bigr)^2\,\widetilde{T}_{--}(\tilde{x})
  +\frac{c-26}{24\pi}\,\bigl\{f,\,x^-\bigr\}\nn
 &=-\frac{c-26}{12\pi}\,\sqrt{\partial_-f}\,\partial_-^2\,
  \frac{1}{\sqrt{\partial_-f}}\nn
 &=-\,4\beta\,\sqrt{\partial_-f}\,\partial_-^2\,
  \frac{1}{\sqrt{\partial_-f}}.
\end{align}

Table \eq{comp} gives us an interpretation 
that in two dimensions, 
the coefficient of $G^{-1}\,\partial_-^2\,G$ 
is shifted from the classical value $n=0$ 
to $n_{\rm eff}\equiv (2/3)(c-26)$. 
This in turn implies that the radiative corrections from 
matter fields keep the form of the conditions.

In the lightcone quantization, the coordinate $x^+$ 
is regarded as time. 
Thus, defining the effective area of a point $(x^+\!\!=\!0,x^-)$ 
on the time-slice $x^+\!=0$ as 
\ba
 A_{\rm eff}(x^-)\equiv A_{\rm eff}(0)
  \,\bigl[ G(x^+\!\!=\!0,x^-)\bigr]^{n_{\rm eff}}
  = A_{\rm eff}(0) \left(\frac{\partial_-f(x^+\!\!=\!0,0)}
  {\partial_-f(x^+\!\!=\!0,x^-)} \right)^{\!(c-26)/3} 
\ea
with unknown constant $A_{\rm eff}(0)$, 
and following the derivation of the Bousso bound given in the preceding 
section with an appropriate initial condition, 
we expect that the entropy on the lightsheet $L$ ($0\leq x^-\leq 1$) 
at time $x^+\!=0$ would be bounded as 
\begin{align}
 S_L&\equiv A_{\rm eff}(0) \int_0^1\,dx^-\,s(x^+\!\!=\!0,x^-)
  \,\bigl[G(x^+\!\!=\!0,x^-) \bigr]^{n_{\rm eff}} \nn
 &\leq -\frac{1}{4}\,A_{\rm eff}(0)\,
  \Bigl[\bigl[G(x^+\!\!=\!0,x^-)\bigr]^{n_{\rm eff}}
  \Bigr]_0^1 \nn
 &=\frac{1}{4}\,\bigl(A_{\rm eff}(0)-A_{\rm eff}(1)\bigr).
\end{align}

%%%%%%%%%%%%%%%%%%%%%%%%%%%%%%%%%%%%%%%%%%%%%%%%%%%%%%%%%%%%%%%%%%%
\resection{2D dilatonic gravity}

In this section, we discuss two-dimensional dilatonic gravity, 
which can be obtained by compactifying an $(n+2)$-dimensional space-time 
on $n$-dimensional sphere, $S^n$.
We demonstrate that $n_{\rm eff}=(2/3)(c-26)$ can be naturally identified 
with the dimensionality of the compactified  space, $n_{\rm eff}\sim n$. 
Our discussion was inspired by work of Strominger and Thompson 
\cite{Strominger:2003br}.

We consider an $(n+2)$-dimensional space-time $M_{n+2}$ 
with topology $M_{n+2}=M_2\times S^n$ and with 
coordinates $X^M=(x^\mu,\,y^i)$ ($\mu=0,1$ (or $+,-$) and $i=1,\cdots,n$). 
The metric is then written as 
\begin{align}
 ds_{n+2}^2&=G_{M\!N}\,dX^M dX^N \nn
 &=g_{\mu\nu}(x)\,dx^\mu dx^\nu + e^{-2\phi(x)}\,\tilde{g}_{ij}(y) dy^i\,dy^j,
\end{align}
where $d\tilde{s}_{n}^2\equiv \tilde{g}_{ij}(y)\,dy^i dy^j$ 
is a metric of unit sphere, 
which can be taken, for example, to be 
$\tilde{g}_{ij}(y)=\delta_{ij}+y_i y_j/(1-y^2)$ with $|y|^2\leq 1$. 
If we take only the zero mode of harmonic functions on $S^n$, 
the Einstein-Hilbert action reduces to the action of dilatonic gravity:
\begin{align}
 S_{n+2}\bigl[G_{M\!N}(x,y)\bigr]
  &=\frac{1}{16\pi G_{n+2}}\,\int\,d^2 x\,d^n y\,
  \sqrt{-G}\,R_G\nn
 &\rightarrow
  \frac{1}{16\pi G_2}\,
  \int\,d^2x\,\sqrt{-g}\,e^{-n\phi}\,
  \Bigl[ R+n(n-1)\Bigl(\bigl(\nabla\phi\bigr)^2+e^{2\phi}\Bigr)\Bigr]\nn
 &\equiv S^{\rm DG}\bigl[g_{\mu\nu}(x),\phi(x)\bigr].
\end{align}
Here $G_2\equiv G_{n+2}/\omega_n$ is the two-dimensional Newton constant 
($\omega_n=\int\,d^ny\,\sqrt{\tilde{g}}$ is the volume of unit sphere) 
and will be set to unity in the following discussion.

We choose the metric $g_{\mu\nu}(x)$ as in the preceding section, 
\ba
 ds_2^2=g_{\mu\nu}(x)\,dx^\mu dx^\nu
  =-dx^+\bigl( dx^-\!+h(x^+,x^-)\,dx^+\bigr).
\ea
Then the vectors, $k\equiv\partial_-$, $l\equiv\partial_+-h\,\partial_-$ 
and $\eta_{(i)}\equiv\partial_{y^i}$, satisfy the following equations:
\begin{align}
 &k^2=l^2=0,\quad k\cdot l=-1/2,\quad
  k\cdot \eta_{(i)}=l\cdot \eta_{(i)}=0\nn
 &k\cdot\nabla\,k^M=0,\quad k\cdot\nabla\,l^M=0,\quad
 k\cdot\nabla\,\eta^M_{(i)}=\hat{B}^M_{~\,N}\,\eta^N_{(i)}
\end{align}
with
\ba
 \hat{B}^M_{~\,N}=-\bigl(\partial_-\phi\bigr)\,
  \Bigl(\delta^M_{~N}+2\,k^Ml_N+2\,l^Mk_N\Bigr).
\ea
{}From this, we find that $x^-$ is an affine parameter 
of a null hypersurface $x^+={\rm const}$, 
and that the expansion $\theta\equiv \hat{B}^M_{~\,M}$ 
is given by
\ba
 \theta=-n\,\partial_-\phi.
\ea
This can also be concluded by noting that the cross-sectional area 
at $x^-$ is given by $A=\omega_n\,e^{-n\phi}$ 
so that the expansion is given as $\theta=\partial_-A/A=-n\,\partial_-\phi$.

The null component of the energy-momentum tensor is calculated as
\begin{align}
 T^{\rm DG}_{--}&\equiv T^{\rm DG}_{\mu\nu}\,k^\mu k^\nu\nn
 &=\frac{2}{\sqrt{-g}}\,\frac{\delta S^{\rm DG}}{\delta g_{\mu\nu}}\,
  k_\mu k_\nu\nn
 &=-\frac{n}{8\pi}\,e^{-n\phi}\,e^\phi\,\partial_-^2\,e^{-\phi}. 
\end{align}
This expression suggests that the null component of the energy-momentum tensor 
of dilatonic gravity is related to that of the preceding section, $T_{--}$, as 
\ba
 T^{\rm DG}_{--}=e^{-n\phi}\,T_{--},
 \label{EM}
\ea
if we identify the quantities there as 
\ba
 e^{-\phi}\sim G=\frac{1}{\sqrt{\partial_-f}}
\ea
and 
\ba
 n_{\rm eff}\sim n. 
\ea

\begin{figure}[htbp]
\begin{center}
\resizebox{!}{60mm}{
 \input{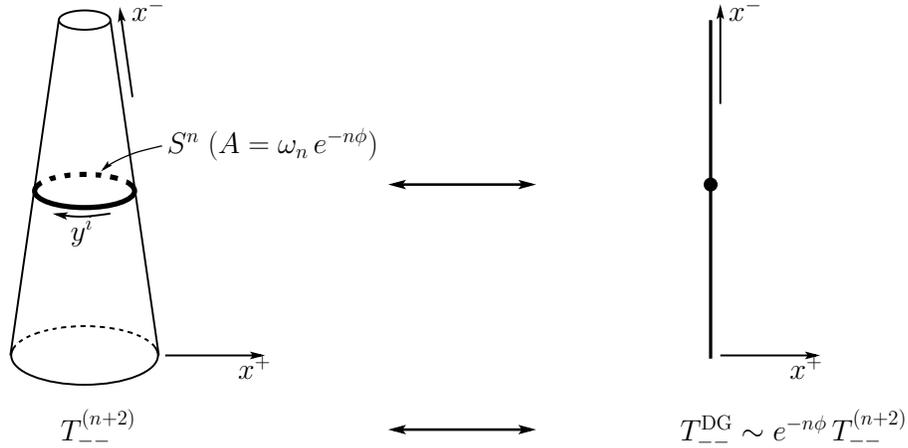}
}
\caption{\footnotesize{ 
Each sphere $S^n$ on the lightsheet of the left figure 
corresponds to a point on the lightsheet %($x^+={\rm const}$) 
of the right figure after the compactification is made 
from $(n+2)$-dimensional space-time to two-dimensional one.
2D energy momentum tensor $T_{--}^{\rm DG}$ thus corresponds 
to the $(n+2)$D energy momentum tensor $T_{--}^{(n+2)}$ 
if we multiply the latter by the cross-sectional area at $x^-$, 
$T_{--}^{\rm DG}=\bigl(\omega_n\, e^{-n \phi}\bigr)\times T_{--}^{(n+2)}$.}}
\label{reduce}
\end{center}
\end{figure}

This identification can be inferred from the observation 
made in ref.\ \cite{Strominger:2003br}, 
that quantities which are scalar on an $n$-dimensional submanifold 
of the lightsheet with fixed affine parameter $x^-$ 
(like entropy density, $s^{(n+2)}$, and 
the null component of the energy-momentum tensor, $T^{(n+2)}_{--}$) 
should be multiplied by the area of the submanifold at $x^-$ 
in order to interpret them as quantities in two-dimensional dilatonic
gravity
(see fig.\ \ref{reduce}): 
\ba
 s^{\rm DG}\sim e^{-n\phi}\,s^{(n+2)},\quad
 T^{\rm DG}_{--}\sim e^{-n\phi}\,T^{(n+2)}_{--}. 
\ea
This implies that the two-dimensional objects like 
the entropy current and the energy-momentum tensor 
in the preceding section are directly related to 
$(n+2)$-dimensional objects through the relation $n_{\rm eff}\sim n$. 
The relation $n_{\rm eff}=(2/3)(c-26)\sim n$ is actually plausible, 
since, if we start with larger $n$, then two-dimensional 
gravity should feel conformal matters of larger central charge. 

We can give another reasoning to the relation \eq{EM} 
that is purely based on two-dimensional consideration. 
We first note that in dilatonic gravity, 
the propagator of the field $h$ is given by 
$\bigl\langle h(p)\,h(-p)\bigr\rangle \sim e^{n\phi}$. 
We further note that $S^{\rm DG}[h,\phi]$ can be interpreted 
as the classical part of the generating functional 
of amputated 1PI diagrams with vacuum expectation values $h(x)$ and $\phi(x)$, 
while $\Gammaeff[h]$ is interpreted as the generating functional 
with $h(x)$ as the source of the energy-momentum tensor. 
This consideration leads to the desired relation 
\ba
 T_{--}\sim \frac{\delta\Gammaeff}{\delta h}
  \sim \bigl\langle h\,h\bigr\rangle\,\frac{\delta S^{\rm DG}}{\delta h}
  \sim e^{n\phi}\,T^{\rm DG}_{--}. 
\ea

%%%%%%%%%%%%%%%%%%%%%%%%%%%%%%%%%%%%%%%%%%%%%%%%%%%%%%%%%%%%%%%%%%
\resection{Conclusion and outlook}

In this letter, we considered two-dimensional gravity, 
treating metric as a background for quantum conformal matters, 
and argued that a nontrivial holographic bound on entropy 
holds even for two-dimensional case due to the Weyl anomaly. 
We also discussed dilatonic gravity to demonstrate 
the naturalness of our relation $n_{\rm eff}\sim (2/3)(c-26)$. 

We further need to quantize the metric $h$ 
in order to see how Bousso's entropy bound is realized 
in full quantum gravity in two dimensions. 
The quantization would be carried out 
with the use of representation theory of $SL(2,\bR)$ Kac-Moody algebra, 
which is the residual gauge symmetry after the lightcone gauge 
is taken \cite{Polyakov:1987zb}\cite{Knizhnik:ak}. 
Work towards this direction is now in progress. 

%%%%%%%%%%%%%%%%%%%%%%%%%%%%%%%%%%%%%%%%%%%%%%%%%%%%%%%%%%%%%%%%%%%

\section*{\large{Acknowledgment}}

The authors would like to thank W.~Hikida, H.~Kawai, M.~Ninomiya 
and T.~Tada for discussions.  

%%%%%%%%%%%%%%%%%%%%%%%%%%%%%%%
%%% reference
%%%%%%%%%%%%%%%%%%%%%%%%%%%%%%%
\baselineskip=0.7\normalbaselineskip

\end{document}